\documentclass[twocolumn,showpacs,preprintnumbers,amsmath,amssymb]{revtex4}

\usepackage{graphicx}
\usepackage{dcolumn}
\usepackage{bm}
\usepackage{color}
\newcommand{\figwidth}{3.375in}

\begin{document}


\title{Phase transition of Two-timescale Two-temperature Spin-lattice Gas Model}

\author{C.~H.~Nakajima}
\email{cnakajima@huku.c.u-tokyo.ac.jp}
\author{K.~Hukushima}
 \email{hukusima@phys.c.u-tokyo.ac.jp}
\affiliation{Department of basic Science, Graduate School of Arts and
 Sciences, University of Tokyo, 3-8-1 Komaba, Tokyo 153-8902, Japan
}

\date{\today}

\begin{abstract}
We study phase transition of a nonequilibrium statistical-mechanical
 model, in which two degrees of freedom
 with different time scales separated from each other touch to their own
 heat bath. A general condition for finding anomalous negative latent
 heat recently discovered is derived a from thermodynamic argument. As a
 specific example, phase diagram of a spin-lattice gas model is studied
 based on a mean-field analysis with replica method. While
 configurational variables are spin and particle in this model, it is
 found that the negative latent heat appears in a parameter region of
 the model, irrespective of the order of their time scale. Qualitative
 differences in the phase diagram are also discussed. 

\end{abstract}

\maketitle

\section{introduction
}

Phase transitions under non-equilibrium conditions have attracted
with a great deal of attention in statistical-mechanical
problems\cite{NEPT,NEPT2,NEPT3}.  
There have been many investigations on non-equilibrium phase transitions
so far\cite{DRIVEN,SHEAR,R,ZPM}, which have revealed a fascinating new
transition behavior different from equilibrium transition. In general, probability
distribution in non-equilibrium cannot be expressed in terms of only
energy functional, which causes a difficulty in theoretical
study. 

Recently  another class of non-equilibrium systems that exhibits
a phase transition has been studied\cite{AP},  in which two different
degrees of freedom coupled to their own heat baths interact with each
other through multi-body interactions. For simplicity, time scales of
these two variables are assumed to be well separated. Then, the systems
consist of slow and fast variables belonging to different time hierarchy.
The fast variables behave in quasi-equilibrium for a given set of slow
variables that plays a role as quenched variables for the fast ones. Meanwhile, the slow variables are
not given by independent distribution function as in quenched disordered
systems but affected through mean force of fluctuating fast variables.  
Such systems with a hierarchy in seperated time scales and
different heat baths are here called ``two-temperature'' system.
These systems are adopted to describe neural network systems with
a synaptic evolution\cite{CPS}, evolving networks\cite{WS}, and some
kind of NMR systems\cite{DFK}. In contrast to most of non-equilibrium
systems, the steady-state distribution of the models is formally
expressed in terms of the energy function using the replica method that
is a standard tool for studying thermodynamic properties of the quenched
disordered systems\cite{MPV}.

The replica formalism for two-temperature systems has been introduced in Refs.~[\onlinecite{CPS, Dotsenko,ANS}]. 
While the quenched disordered systems require the replica limit in which
the replica number takes to be zero, the two-temperature systems have a
physical meaning for any value of the replica number which corresponds
to a ratio of two temperatures. In this sense, this is regarded as a
generalization of quenched systems and is sometimes called as partial
annealing system\cite{CPS}. A most crutial difference from the quenched
systems is that the slow variables are also dynamical coupled to their
heat bath. Therefore, both the slow and fast variables are responsible
for phase transition. This could provide a new cooperative phenomenon
over different time scale. 

In fact,  Allahverdyan and Petrosyan, hereafter referred to as AP,
studied a mean-field spin  model as a two-temperature system
and found that the model exhibited a first-order phase transition with
anomalous negative latent heat, which never occurs in equilibrium
statistical mechanics. However, this peculiar behavior observed in
the two-temperature system is not well understood. We pursue phase
transition in the two-temperature systems and give a general condition
that the system exhibits the negative latent heat with the help of the
idea of thermodynamics.  It is also found that, in the systems, two
different entropys associated with the fast and slow variables
respectively play a competitive role in determing the phase boundary of
first-order transition. We further studied two-temprature version of a
spin-lattice-gas model, similar to that studied by AP,  as a specific
example. The spin-lattice-gas model consists of two degrees of freedom,
spins and particles, which has been studied for a given Hamiltonian in
equilibrium\cite{Sok}.   The two-temperature version is
characterized by not only the Hamiltonian but also the order of time
scales of two variables. AP studied the case where the spins were slow
and the particles were fast. We study this case with some modified
Hamiltonian and also the other case, namely the spins and particles
behave as fast and slow variables, respectively. We then find that the
existence of the negative latent heat is common to both cases,
suggesting that it is observed in a wide class of the two-temperature
systems. On the other hand, qualitatively different behavior is also
found in phase diagram of the two cases, in particluar, the stability of
ferromagnetic ordered phase. 

This paper is organized as follows.
In Sec.~\ref{sec:formulation}, we review the replica formalism of the
two-temperature system, which leads to an equilibrium model of a
replicated system. 
We also discuss  phase boundary of first-order transition and derive a 
Clausius-Clapeyron relation in this system. 
This relation enables us to find generally a geometric property of the
phase boundary and the negative latent heat.
In Sec.~\ref{sec:mfmodel}, we explicitly define two mean-field
spin-lattice-gas models with different order of time scales  and give
self-consistent equations for the models.   
The results obtained by solving the equations are presented in  
Sec.~\ref{sec:results}. 
In Sec.~\ref{sec:summary}, we summarize our results.

\section{Two-temperature formalism with different time scales}
\label{sec:formulation}
In this section, we review a theoretical formalism for two-time-scale and two-temperature system\cite{AP}.
Suppose a system described by a Hamiltonian $H(f,s)$, in which $f$ is a
symbolic notation of fast degree of freedom and $s$ is of slow degree
of freedom.  These variables $f$ and $s$ are in contact with their
different heat bath with temperature $T_f$ and $T_s$, respectively.
We assume that the two characteristic time scales on the variables $s$
and $f$ are well separated from each other and that the thermal average
of the fast variable $f$ can be taken with a fixed configuration of the
slow variable $s$. Then, the conditional probability $P(f|s)$ of finding
a configuration $f$ for a given $s$ at the inverse temperature
$\beta_f=1/T_f$ is defined
\begin{equation}
P(f|s) = \frac{e^{-\beta_f H(f,s)}}{Z(s)},
\label{eq:fastdistr}
\end{equation}
where the normalization constant or the partition function of the fast
variable is set as
\begin{equation}
Z(s) = \mathrm{Tr}_f e^{-\beta_f H(f,s)}. 
\label{eq:fastpf}
\end{equation}
Hereafter the Boltzmann constant is set to be unity.
One can define partial free energy for the fast variable as $F_f(s) = -T_f
\log {Z}(s)$.

The steady-state probability of slow variables $P(s)$ is derived by an
adiabatic approximation of two-temperature Langevin equation
\cite{ANS}.
The force acting on $s$ is assumed to be an averaged derivative of the
Hamiltonian with respect to the slow variable over the conditional
probability, which is represented by the partial free energy as
$-\frac{\partial F_f(s)}{\partial s}$.
The equilibrium distribution $P(s)$ at the inverse temperature $\beta_s=1/T_s$ is given by
\begin{eqnarray}
P(s) &=& \frac{e^{-\beta_s F_f(s)}}{\mathcal{Z}} \label{eq:slowdistr}
\end{eqnarray}
where
\begin{eqnarray}
\mathcal{Z} &=& \mathrm{Tr}_s e^{-\beta_s F_f(s)}.  \label{eq:slowpf}
\end{eqnarray}
The total free energy $\mathcal{F}$ is defined by $\mathcal{F}=-T_s \log
{\mathcal{Z}}$.

Using the replica trick, the model can be mapped onto an equilibrium problem with a replicated Hamiltonian for the integer number of ratio $n=T_f/T_s$,
\begin{eqnarray}
\mathcal{F} &=& -T_s \log \left(\mathrm{Tr}_s \mathrm{Tr}_f^{(1)} \cdots \mathrm{Tr}_f^{(n)} e^{-\beta_f \sum_{l=1}^n H(f^{(l)},s)}\right) \label{eq:replicafe} ,
\end{eqnarray}
where $f^{(l)}$ denotes replicated fast variables.
This could be extended to any real value of the ratio $T_f/T_s$ after calculating the
replicated system in a standard manner of the replica method.
While one takes the replica limit $n\rightarrow 0$ for the quenched
disordered system like spin glasses, any value of $n$ makes a sense as
the two-temperature system in this context.
This formalism is also interpreted as a kind of statistical-mechanical problems with randomness.
In particular, note that the distribution of the random
variables is determined by not only a given independent function but also the
thermal averaged quantity of the fast variables. The latter leads to a non-trivial correlation among the slow variables.

We discuss thermodynamic properties of the two-temperature system.
The simultaneous probability $P(f,s)$  is expressed as
$P(f,s)=P(f|s)P(s)$.
The total entropy $S$ defined by the simultaneous probability is
decomposed into two degrees of freedom  as
\begin{equation}
 S = -\mathrm{Tr}_{s,f}P(f,s)\log P(f,s) = \mathcal{S}_s +S_f,
\end{equation}
where $S_s$ and $S_f$ are expressed as
\begin{eqnarray}
 \mathcal{S}_s & = & -\mathrm{Tr}_sP(s)\log P(s)\label{eq:entropy-one}, \\
S_f & =& -\mathrm{Tr}_sP(s)\left(\mathrm{Tr}_f P(f|s)\log P(f|s)\right)\label{eq:entropy-two}.
\end{eqnarray}
The total free energy $\mathcal{F}$ is formally expressed as
\begin{equation}
 \mathcal{F}(T_s,T_f) = \overline{\langle H(f,s)\rangle_f} -T_f S_f -T_s\mathcal{S}_s,
\end{equation}
where $\langle\cdots\rangle_f$ and $\overline{\cdots}$ denote an average
over the variables $f$ with $P(f|s)$ and $s$ with $P(s)$,
respectively. It should be noted that
the free energy is also rewritten by
\begin{equation}
 \mathcal{F}(T_s,T_f) = \overline{F_f(s)} -T_s\mathcal{S}_s.
\end{equation}
Averaging over the fast variables $f$, the thermodynamic structure is
found by regarding the averaged partial free energy $\overline{F_f(s)}$
as an ``\textit{energy}'' for the
slow variable $s$. Namely, the averaged partial free energy and the
entropy $\mathcal{S}_f$ for the slow variables decrease monotonically with
decreasing $T_s$.

As a consequence of the thermodynamic structure\cite{EF}, 
a Clausius-Clapeyron like relation  for two-temperature systems is 
derived, which gives us a topological property of a first-order-transition line.
Suppose a phase diagram of the system onto
two-temperature plane of $T_s$ and $T_f$.
We take two points $(T_f,T_s)$ and $(T_f+\delta T_f, T_s+\delta T_s)$
which are located on either side of the first-order-transition line.
The free-energy difference $\delta F$ between these points with small
temperature differences $\delta T_f$ and $\delta T_s$ is given by
\begin{equation}
 \delta\mathcal{F} = -S_f\delta T_f -\mathcal{S}_s\delta T_s.
\label{eq:deriv-e}
\end{equation}
At the first-order transition point $(T_f^{(1)},T_s^{(1)})$, the ordered
and disordered states coexist and the free energy of these states
coincides with each other, meaning
\begin{equation}
\Delta\mathcal{F} = \mathcal{F}^{(o)}(T_f^{(1)},T_s^{(1)}) - \mathcal{F}^{(d)}(T_f^{(1)},T_s^{(1)})=0,
\label{eq:fotcond-e}
\end{equation}
where the upper suffixes $o$ and $d$ denote the ordered and the disordered
states, respectively,  and $\Delta A$ means the difference of a physical
quantity $A$ between the ordered and disordered states at the
transition point.
Using Eqs.~(\ref{eq:deriv-e}) and (\ref{eq:fotcond-e}), the
Clausius-Clapeyron\cite{EF} like relation is obtained as
\begin{equation}
\frac{\delta T_s^{(1)}}{\delta T_f^{(1)}} = - \frac{\Delta S_f}{\Delta \mathcal{S}_s}.
\label{eq:cc-e}
\end{equation}
This implies that 
when the slope of the phase boudary $dT_s^{(1)}/dT_f^{(1)}$ is positive, 
$\Delta \mathcal{S}_s$ and $\Delta S_f$ are opposite to each other.
The free-energy difference $\Delta\mathcal{F}$ is
also expressed as $\Delta\mathcal{F}=\Delta\mathcal{U}-T_f\Delta
S_f-T_s\Delta \mathcal{S}_s$
with $\mathcal{U}$ being the internal energy $\overline{\langle
H\rangle_f}$. Thus, we obtain the relation between the deference of the
internal energy
and the phase boundary as
\begin{equation}
\Delta\mathcal{U}(T_f^{(1)}, T_s^{(1)}) = T_s^{(1)}\Delta \mathcal{S}_s\left(1-n\frac{dT_s^{(1)}}{dT_f^{(1)}}\right).\label{eq:TT-CC-form2}
\end{equation}
Because the entropy $\mathcal{S}_s$ is a monotonically decreasing function of
$T_s$, the sign of $\Delta\mathcal{U}$ depends on only the
gradient of the phase boundary.
This implies that the condition to find the negative latent heat is $\frac{1}{n} \le dT_s^{(1)}/dT_f^{(1)}$ 
when $T_s$ decreases.
On the other hand, when $T_f$ decreases, the condition for the negative latent heat changes to 
$0 \le dT_s^{(1)}/dT_f^{(1)} \le \frac{1}{n}$.
While AP explicitly found that a specific spin-lattice gas model
exhibited the negative latent heat in a region of the phase diagram using
the replica method, we find a general condition for which the
negative latent heat appears thorough the thermodynamic argument.

\section{Mean-field spin-lattice gas model}
\label{sec:mfmodel}
A model Hamiltonian we studied is an infinite-range spin-lattice-gas
model, that is given by
\begin{equation}
H(\{S_i,p_i\}) = -\frac{1}{N} \sum_{(ij)} \Big( J S_i S_j +\epsilon_f
 \Big) p_i p_j + \alpha \sum_{i=1}^Np_i
\label{eq:hamiltonian}
\end{equation}
where $S_i=\pm1$ are spin variables, $p_i=0,1$ are particle occupation
variables and they are defined on $N$ sites.
In the case  where the spins $S_i$ are the slow
variable and the particles $\rho_i$ the fast, referred
to as case-1 model, the model system with $\epsilon_f=0$ is identical with 
that studied by AP\cite{AP}. We also consider the inverse case where the
spins and the particles represent the fast and slow variables,
respectively, which is referred to as case-2 model.
The spin and particle variables are coupled to their own
heat baths whose temperature is denoted by $T_S$ and $T_p$,
respectively.
The sum is taken over all pairs of sites. The interactions $J$  and
$\epsilon_f$  denote a ferromagnetic coupling and an attractive
interaction between particles, respectively. 
In this paper, $J$ is taken as a unit of energy and
temperature.
The first term of the Hamiltonian consists of a spin mediated
interaction and direct one.
The second term plays a role for controlling a particle number with chemical
potential $\alpha$ which is chosen to be a positive value.
The spin-lattice-gas model could exhibit two types of phase transition
which are  magnetic and density orderings.
The interaction $-\Big( J S_i S_j +\epsilon_f \Big)$ between particles
prefers to increase the particle density and magnetically ferromagnetic 
ordering, while the chemical potential $\alpha$ tends to decrease the 
particle density.
In this sense, these two energy terms compete with each
other. Furthermore, two different kinds of the entropy
associated with the fast and slow variables also compete with the energy
terms.

Since the Hamiltonian is an infinite range model, the trace of
Eq.~(\ref{eq:replicafe}) is carried out with the help of the replica method
by introducing two auxiliary fields $m$ and $\rho$,
which correspond to average magnetization and particle density, respectively.
The self-consistent equations for $m$ and $\rho$ are written as, for the
case-1 model,
\begin{eqnarray}
 m &=& \frac{\displaystyle\sum_{S=\pm 1}S\phi(S;m,\rho)\left(
1+\phi(S;m,\rho)\right)^{n-1}}{\displaystyle\sum_{S=\pm 1}
\left( 1+\phi(S;m,\rho)\right)^n} \label{eq:m1}, \\
 \rho &=& \frac{\displaystyle\sum_{S=\pm 1}\phi(S;m,\rho) \left( 1+\phi(S;m,\rho)
\right)^{n-1}}{\displaystyle\sum_{S=\pm 1} \left( 1+\phi(S;m,\rho) \right)^n}  \label{eq:r1},
\end{eqnarray}
and for the case-2 model,
\begin{eqnarray}
m &=&\frac{ \Big( \displaystyle\sum_{S=\pm 1}S\phi(S;m,\rho) \Big) \Big( \displaystyle\sum_{S=\pm 1}\phi(S;m,\rho)  \Big)^{n-1} }
{ 2^n + \Big( \displaystyle\sum_{S=\pm 1}\phi(S;m,\rho) \Big)^n } \label{eq:m2},\\
\rho &=& \frac{\Big( \displaystyle\sum_{S=\pm 1}\phi(S;m,\rho) \Big)^n }
{ 2^n + \Big( \displaystyle\sum_{S=\pm 1}\phi(S;m,\rho) \Big)^n } \label{eq:r2} ,
\end{eqnarray}
where $\phi(S;m,\rho) = e^{-\beta_f \Big( \alpha - mJS-\epsilon_f \rho \Big) }$.
Here $\beta_f$ is the inverse of the fast
temperature, which corresponds to $1/T_p$ in the case-1 model 
and $1/T_S$ in the case-2 model.
The free energy of system is represented with a solution $(m_0,\rho_0)$ of the 
above self-consistent equations as
\begin{eqnarray*}
\mathcal{F}_{1}(m_0,\rho_0,T_S,T_p)&=&\frac{1}{2}\Big(Jm_0^2+\epsilon_f\rho_0^2\Big)
\end{eqnarray*}
\begin{eqnarray}
\qquad\qquad
-T_S\log\left(\sum_{S=\pm 1}
\left(1+\phi(S;m_0,\rho_0)\right)^n
\right)
\end{eqnarray}
for the case 1 and
\begin{eqnarray*}
\mathcal{F}_{2}(m_0,\rho_0,T_S,T_p)=\frac{1}{2}\Big(Jm_0^2+\epsilon_f\rho_0^2\Big)
\end{eqnarray*}
\begin{eqnarray}
\qquad\qquad 
-T_{p}\log\Bigg(2^n + ( \ \sum_{S=\pm1}\phi(S;m_0,\rho_0) \ )^n\Bigg)
\end{eqnarray}
for the case-2. Here $n$ is defined as the ratio of the fast temperature to
the slow one, namely $n$ is $\frac{T_p}{T_S}$ and $\frac{T_S}{T_p}$ for
the case-1 and case-2, respectively. In this paper,
$\mathcal{F}_1$ and $\mathcal{F}_2$ as functions of $m$ and $\rho$ are loosely called  to
``free energy'' in a sense of Ginzburg-Landau free energy.
According to the Eqs.~(\ref{eq:entropy-one}) and (\ref{eq:entropy-two}), two kinds of decomposed entropy 
are termed as $S_S$ and $\mathcal{S}_p$, respectively for the case-1 model, and $S_p$ and $\mathcal{S}_S$ for the case-2 model.

\section{Results and discussions}
\label{sec:results}
\subsection{Phase diagram and negative latent heat}
\begin{figure}
\includegraphics[width=\figwidth]{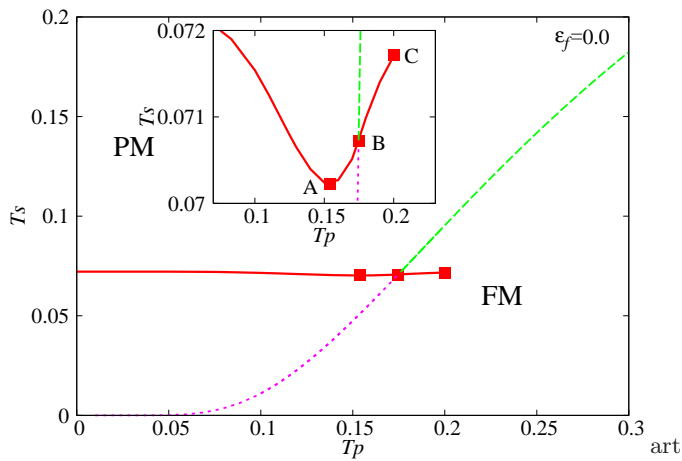}
art
\caption{
Phase diagram of the case-2 model on $T_S-T_p$ plane  with $\alpha=0.45$
 and $\epsilon_f=0$. PM and FM denote paramagnetic and
 ferromagnetic phases, respectively. The solid and dashed lines represent
 the first and  second order transitions and the dotted one represents the
 instability limit of the paramagnetic solution.
The inset shows an enlarged view around the critical point.
}
\label{fig:rap0pd}
\end{figure}
\begin{figure}
\includegraphics[width=\figwidth]{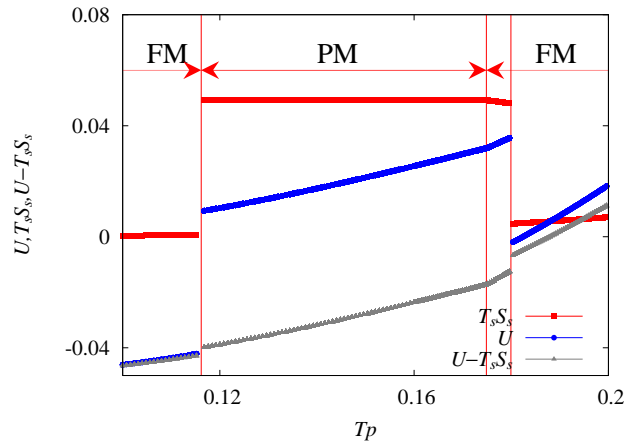}
\caption{
$T_p$ dependence of $\mathcal{U}$, $T_S\mathcal{S}_{S}$, and
averaged partial free energy $\overline{F_S(p)}=\mathcal{U}-T_S\mathcal{S}_{S}$ at
 $\alpha=0.45$ and $T_S=0.071$ in the case-2 model.
There are three phase transitions, indicated by vertical lines, with
 $T_p$ changing for a fixed $T_S$. Two first-order transition occurs at
 $T_p=0.116$   and $0.180$. The former is between PM and FM, whereas the
 latter is a re-entrant transition between dense and dilute
 ferromagnetic phases. Between these transitions, a second-order
 transition occurs at $T_p=0.175$.
}
\label{fig:rapef-dm071-sum0c}
\end{figure}
We first discuss phase diagram of the spin-lattice gas
model with the condition $\epsilon_f=0$ both for the case-1 and the
case-2 models. The case-1 model with $\epsilon_f=0$, that is the same as that studied by AP\cite{AP}, shows that 
a ferromagnetic phase has a place in a low $T_S$ region at $\alpha=0.45$, 
and that there is a region of the phase boundary in which the internal energy of the ferromagnetic phase is higher than that of the paramagnetic phase.
Namely,  the phase transition involves the negative latent heat discussed in Sec.~\ref{sec:mfmodel}.

We study phase diagram of the case-2 model, the time scale reversed
version studied by AP\cite{AP}. Figure~\ref{fig:rap0pd} shows 
a phase diagram
on the $T_S-T_p$ plane for the case-2 model with
$\epsilon_f=0$ and $\alpha=0.45$,  which is in comparison to the phase
diagram of the case-1 model shown in Ref.~\cite{AP} under the condition
of $\epsilon_f=0$.

While the first-order phase-transition temperature shows rather weak dependence of
$T_p$ and takes a finite value at $T_p=0$,  it behaves
non-monotonically as a function of $T_p$ near the {critical}  point as shown in the inset of Fig.~\ref{fig:rap0pd}.
According to Eq.~(\ref{eq:TT-CC-form2}), in the region between A and B shown in the figure, the latent heat becomes anomalously negative when $T_p$ decreases.
Fig.~\ref{fig:rapef-dm071-sum0c} shows $T_p$ dependence of thermodynamic
quantities for a fixed $T_S$, where phase
transitions occur three times as a function of $T_p$.
As $T_p$ decreases at $T_S=0.071$,
a first-order transition occurs at $T_p=0.18$ from a dense ferromagnetic
phase to a dilute one and a second-order transition between the dilute
ferromagnetic and the paramagnetic phases at $T_p=0.175$. Eventually,
the transition from the paramagnetic to the ferromagnetic phases again
occurs $T_p=0.116$

At the highest transition temperature $T_p=0.180$,
the internal energy and the entropy $S_S$ have a positive jump, while the
averaged partial free energy $\overline{F_S(p)}$ decreases monotonically.
This means that
the internal energy of highly ordered phase at higher temperatures is
lower than that of disordered phase at lower temperatures.
A similar first-order transition is found between A and B in
Fig.~\ref{fig:rap0pd}.
Thus, this phase transition could be simply interpreted as a kind of
reentrant transition, in which the low-temperature disordered phase is
stabilized by an entropic effect.
This is in contrast to the case-1 model where the low-temperature phase is the disordered paramagnetic one.

Another difference between the case-1 and case-2 models is 
found in the topology of the phase diagram.
Whereas the case-1 model has a tricritical point at which the first and the
second-order-transition lines merge, the first-order-transition line
enters into the ferromagnetic phase in the case-2 model as shown in
Fig.~\ref{fig:rap0pd}.
Interestingly, a density origin phase transition occurs in the ferromagnetic phase.
Near the transition the free-energy has four different local minima
which correspond to high-density and low-density ferromagnetic states
and their time-reversal ones.
This is qualitatively different from that observed by AP in the case-1 model. 

\begin{figure}
\includegraphics[width=\figwidth]{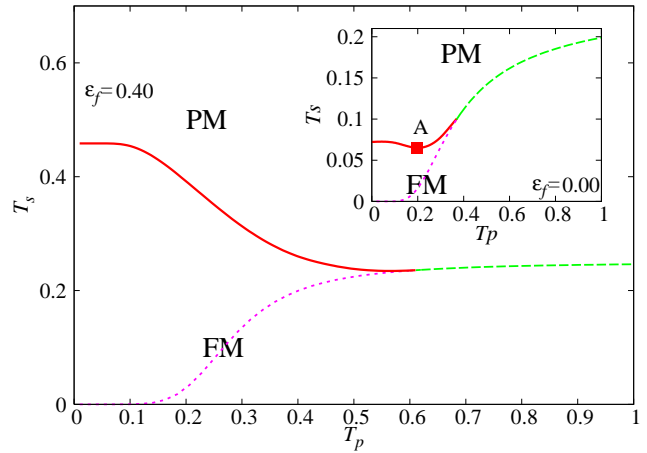}
\caption{
Phase diagram of the case-1 model on $T_S-T_p$ plane with $\alpha=0.45$
 and $\epsilon_f=0.4$. The thick line represents the first-order phase
 transition. 
The symbols of lines are the same as those in Fig.1
The inset shows the phase diagram of the case-1 model with $\alpha=0.45$ 
and $\epsilon_f=0$.
}
\label{fig:apef-pb-fp-40}
\end{figure}

Let us discuss the effect of $\epsilon_f$ term in the spin-lattice gas
model.
The first-order transition of this system is originated with the
particle density.
Therefore, the first-order-transition line could be changed by introducing the direct interaction between particles,
the $\epsilon_f$ term in Eq.~(\ref{eq:hamiltonian}). We study how the effect of the $\epsilon_f$ term on the phase diagram of both the case-1 and case-2 model.
First, we focus on $\epsilon_f$-dependence of the region in which the
negative latent heat is observed.
Fig.~\ref{fig:apef-pb-fp-40} shows the phase diagram with
$\epsilon_f=0.4$ in the case-1 model and the inset shows that with $\epsilon_f=0$.
As the value of $\epsilon_f$ increases from zero, the first-order
transition temperature $T_{S}^{(1)}$ for a fixed $T_p$ increases and the
ferromagnetic region is extended.
Non-monotonic behavior of $T_S^{(1)}$ found in the inset of
Fig.~\ref{fig:apef-pb-fp-40} near the multicritical point disappears
with $\epsilon_f$ increasing.
Eventually, at the value $\epsilon_f=0.40$ as shown in
Fig.~\ref{fig:apef-pb-fp-40}, the first-order-transition line is
monotonic as a function of $T_p$.
The argument in Sec.~\ref{sec:formulation} yields that the monotonic behavior of $T_S^{(1)}$ as a function of $T_p$ means the absence of negative latent heat on the transition.
Thus, it is found that the region in which negative latent heat observed is robust against an infinitesimal attractive interaction and disappears by further increasing the interaction.
This suggests that the negative latent heat is not peculiar behavior in the two-temperature system and could be observed by controlling the model parameter.
Similar behavior is observed in the case-2 model.
The phase diagram with $\epsilon_f=0.4$ for the case-2 model is shown in
Fig.~\ref{fig:rapef-pb-fp-40}. 
As seen in the case-1 model, the ferromagnetic phase transition is also
enhanced and the non-monotonic region of the first-order-transition line
becomes narrow with $\epsilon_f$ increasing.

\subsection{Stability of ferromagnetism in the two models}

\begin{figure}
\includegraphics[width=\figwidth]{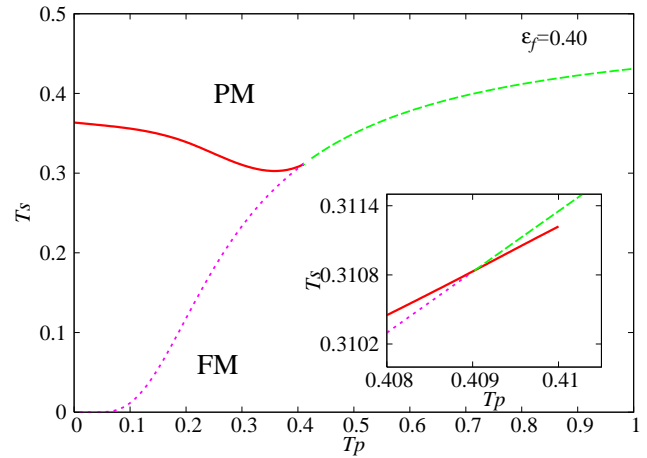}
\caption{
Phase diagram of the case-2 model with $\alpha=0.45$ and
 $\epsilon_f=0.40$.
The symbols of lines are the same as those in Fig.~\ref{fig:rap0pd}.
The inset shows an enlarged view near the multicritical point. The
 first-order-transition line intersects with the paramagnetic
 instability line.
}
\label{fig:rapef-pb-fp-40}
\end{figure}

\begin{figure}
\includegraphics[width=\figwidth]{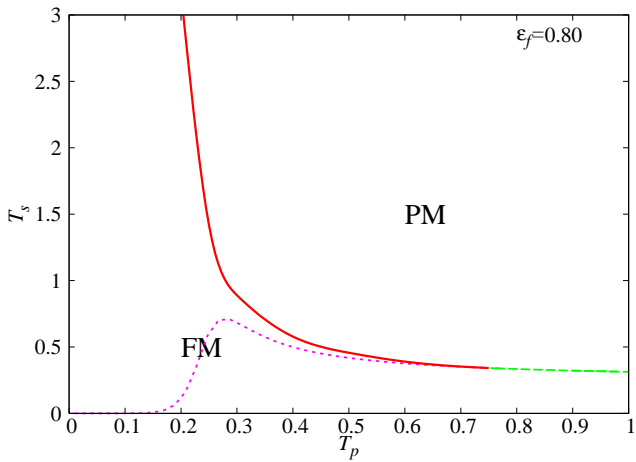}
\caption{
Phase diagram of the case-1 model with $\alpha=0.45$ and $\epsilon_f=0.80$.
The symbols of lines are the same as those in Fig.~\ref{fig:rap0pd}.
}
\label{fig:apef-pb-fp-80}
\end{figure}

\begin{figure}
\includegraphics[width=\figwidth]{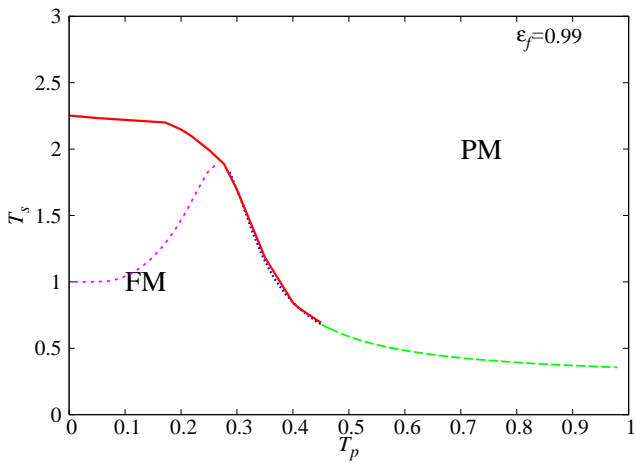}
\caption{
Phase diagram of the case-1 model $\alpha=0.45$ and $\epsilon_f=0.99$.
The symbols of lines are the same as those in Fig.~\ref{fig:rap0pd}.
}
\label{fig:apef-pb-fp-99}
\end{figure}

In this subsection, we discuss phase diagram with relatively large $\epsilon_f$.
There is remarkable difference in the stability of ferromagnetic order between the case-1 and case-2 models.
Fig.~\ref{fig:apef-pb-fp-80} shows a phase diagram of the case-1 model
with $\epsilon_f=0.8$, in which the ferromagnetic phase exists stably up
to extremely high temperature.
As $\epsilon_f$ increases further, $T_S^{(1)}$ takes a finite value in
the limit $T_p=0$ again, as shown in Fig.~\ref{fig:apef-pb-fp-99}.
Namely, in a finite range of $\epsilon_f$, $T_S^{(1)}$ diverges as a function of $T_p$ and then the ferromagnetic phase becomes stable even in the high $T_S$ limit.
In contrast, the ferromagnetic phase boundary in the case-2 model
changes modestly with $\epsilon_f$ increasing as shown in
Fig.~\ref{fig:rapef-pb-fp-40} and $T_S^{(1)}$ remains to be finite in
the limit $T_p=0$ .
This suggests that  the difference of the time scales between the particle and the spin strongly affects on the stability of the ferromagnetic phase.

It would be helpful for making clear the issue mentioned above in the two models to see an instability condition of the paramagnetic phase.
The paramagnetic instability line, $(T_S^{(pmi)},T_p^{(pmi)})$, on the
$T_S-T_p$ plane is simply determined by the condition
$\left.\frac{\partial^2 \mathcal{F}}{\partial m^2}\right|_{m=0,\rho=\rho_{pm}}=0$,
because off-diagonal term of a Hessian matrix of the free energy with respect to $m$ and $\rho$ vanishes in the paramagnetic phase.
Then, the self-consistent equations for $\rho$, Eqs.~(\ref{eq:r1}) and
(\ref{eq:r2}), in the paramagnetic phase are simply reduced to the
equation,
\begin{eqnarray}
\rho_0^{(pm)}&=&\frac{e^{\beta_p(\epsilon_f\rho_0^{(pm)}-\alpha)}}{1+e^{\beta_p(\epsilon_f\rho_0^{(pm)}-\alpha)}},
 \label{eq:density-on-pmi}
\end{eqnarray}
where $\rho_0^{(pm)}$ denotes a solution of the equation in the
paramagnetic phase.
By using the solution of the equation, the instability condition for
the case-1 model is given by
\begin{eqnarray}
\beta_SJ\left(1-\frac{J}{T_p}\rho_0^{(pm)}(1-\rho_0^{(pm)})-\frac{J}{T_S}{\rho_0^{(pm)}}^2\right)=0 \label{eq:AP-Hessian}.
\end{eqnarray}
This yields the instability temperature  $T_S^{(pmi)}$  as a function of
$T_p$ expressed as
\begin{eqnarray}
T_S^{(pmi)}(T_p)=\frac{J{\rho_0^{(pm)}}^2}{1-(J/T_p)\rho_0^{(pm)}(1-\rho_0^{(pm)})} \label{eq:AT-pmi}.
\end{eqnarray}
When the denominator $1-(J/T_p)\rho_0^{(pm)}(1-\rho_0^{(pm)})$ is zero, $T_S^{(pmi)}$ diverges and hence the ferromagnetic phase becomes stable
even at $T_S=\infty$. As a trivial example, when $\epsilon_f=2\alpha$,
$T_S^{(pmi)}$ goes to infinity at
$T_p=1/4$. At $(\epsilon_f,\alpha)=(0.90,0.45)$, the
first-order-transition line and the paramagnetic instability line almost
merge and the jump of thermodynamic quantities at first-order transition
is quite weak in a wide region of $(T_S,T_p)$ plane.
Because the instability line is located on the second-order transition
or inside the ferromagnetic phase, the divergence of $T_S^{(pmi)}(T_p)$
means the stability of ferromagnetic phase at an infinite $T_S$.

We show explicitly the stability of ferromagnetic phase in the
case-1 model at $T_p=0$. In the case-1 model, the spins that are slow
variable can fluctuate even at $T_p=0$. The particle configuration is
adaptively determined for a given slow spin configuration by minimizing the free
energy.
For intermediate $\epsilon_f$, which is, to be precise, given by
$\epsilon_f<1.45$ at $\alpha=0.45$, the paramagnetic state is an empty
state at $T_p=0$, namely $m_0=0$ and $\rho_{pm}=0$. On the other hand,
the  self-consistent equations, Eq.(\ref{eq:m1}) and Eq.(\ref{eq:r1}),
for the ferromagnetic solution, $m_0$ and $\rho_0^{(fm)}$,   at $T_p=0$ are then
\begin{eqnarray}
m_0&=&\pm\rho_0^{(fm)}, \\
\rho_0^{(fm)}&=&\frac{e^{\beta_{S}\big((J+\epsilon_f)\rho_0^{(fm)}+\alpha\big)}}{e^{\beta_{S}\big((J+\epsilon_f)\rho_0^{(fm)}+\alpha\big)}+1}.
\end{eqnarray}
When $T_S$ increases to infinity, $\rho_0^{(fm)}$ decreases gradually down to
$1/2$, but never reaches to zero. Consequently, the magnetization $m$
remains finite even at $T_S=\infty$.
In fact, in the limit $T_S\rightarrow\infty$, the free-energy difference
between the ferromagnetic and the paramagnetic solution takes the form
$-\frac{J+\epsilon_f}{8}+\alpha/2$, that is the internal energy for the
ferromagnetic solution. This yields the stability condition of the
ferromagnetic phase as $\epsilon_f>4\alpha-J$.
For example with $\alpha=0.45$ and $\epsilon_f=0.8$, as shown in
Fig.~\ref{fig:apef-pb-fp-80}, the ferromagnetic phase is extended up to
very high $T_S$ temperature, although the instability line of the
paramagnetic solution goes down to the origin.

For sufficiently large $\epsilon_f$, the paramagnetic solution is
qualitatively changed by the effect of the attractive interaction. Then,
$\rho_0^{(pm)}=1$ at $T_S\rightarrow\infty$ in the limit $T_p=0$ and the
free-energy difference is modified to
$-\frac{J+3\epsilon_f}{8}-\frac{\alpha}{2}$. The dense paramagnetic
solution becomes dominant at $(T_p,T_S)=(0,\infty)$. Thus, the
first-order transition temperature $T_S^{(1)}(T_p)$ can diverges
only in a finite range of $\epsilon_f$ in the case-1 model.

In the case-2 model, on the other hand, the spin variables fluctuate as
fast degree of freedom for a given slow particle configuration.
The paramagnetic instability condition is then given by
\begin{eqnarray}
\beta_pJ\left(1-\frac{1}{T_S}J\rho_0^{(pm)}\right)=0,
\end{eqnarray}
where $\rho_0^{(pm)}$ is again determined by Eq.~(\ref{eq:density-on-pmi}).
The $T_p$ dependent term coupled to ${\rho_0^{(pm)}}^2$ cancels out
because of the symmetry of the fast spin variable.
In the paramagnetic phase, the particle density $\rho_0^{(pm)}$ of the
case-2 model is the same value as the case-1 model.
Thus, $T_S^{(pmi)}$ could not diverge in any value of $\epsilon_f$ and
$T_p$, in sharp contrast to the case-1 model.
This is, however, an necessary condition but not the sufficient one for
the finite transition temperature at $T_p=0$.

We see again the phase boundary at $T_p=0$.
In the case-2 model, the only particle configuration that minimizes the
partial free energy at $T_p=0$ contributes to the ensembles
and the fast spin variables fluctuate under the resultant particle
configurations.
The self-consistent equation for $\rho_0^{(pm)}$ leads to $\rho_0^{(pm)}=0$ at
$T_p=0$, while the corresponding equation for the ferromagnetic phase leads
to a fully occupied solution with $\rho_0=1$. For the latter, the
magnetization $m_0$ is determined by the equation
\begin{eqnarray}
m_0=\tanh\beta_{S}Jm_0,   \label{eq:rapef-ztpt-mag}
\end{eqnarray}
under the condition
$e^{\beta_S(\epsilon_f-\alpha)}\cosh\beta_SJm_0>1$.
Thus, $T_S^{(1)}$ never diverges and the ferromagnetic phase only
emerges at most $T_S<1/J$.
Actually, $T_S^{(1)}$ is obtained by solving the equation
\begin{eqnarray}
0=\frac{1}{2}\big(Jm_0^2+\epsilon_f\big)-T_S\log\big(e^{\beta_S(\epsilon_f+\alpha)}\cosh \beta_SJm_0\big),
\end{eqnarray}
which is derived from the condition that the free-energy difference becomes zero at the transition
temperature.

\section{summary}
\label{sec:summary}

We have studied phase transition of a non-equilibrium
statistical-mechanical model that consists of two degrees of freedom
with different time scales and heat baths, called two-temperature systems. A theoretical framework based
on the replica method and its thermodynamical structure, which have been
already given in the literature\cite{CPS,Dotsenko,AP}, are summarized. As
a direct consequence of the structure, we have pointed out the existence
of a Clausius-Clapeyron like relation in two-temperature systems, which
enables us to link the topology of phase diagram and discontinuity of thermodynamic 
quantities at first-order transition.  In particularly, a general
condition to find the anomalous negative latent heat that is found in a
specific spin model\cite{AP} is reduced to a simple topological
constraint on the phase diagram. To be concrete, when the slope of the
first-order phase boundary is a certain value determined by the retio of two temperatures, 
the negative latent heat appears. It should be worth noting that this
criteria can be applied to any model including short-ranged models in finite
dimensions. 

We have also performed a mean-field analysis of two-temperature version
of a spin-lattice gas model that has  spins and particles as
configurational variables. Generally, two-temperature systems are
characterized by the Hamiltonian and time-scale order of two
varibles. Even in the same Hamiltonian, phase diagram still depends on
the choise of the time-scale order. We have studied phase diagram of the
spin-lattice gas model for two different cases; one is that the spins
are slow and the particles are fast, which is the same as that studied
by AP\cite{AP}, and the other is alternative. Furthermore, the effect by
introducing preferentially an attractive interaction for one of the two
variables is studied. We have found that the general condition for the
negative latent heat is satisfied in a parameter region both for two
cases, suggesting that the negative latent heat is not accidental but
frequently observed in two-temperature systems. By increasing the
attractive interaction, the parameter region 
becomes narrow in common. On the other hand, qualitatively different
properties are found in the phase diagram, such as the stability of the
ferromagnetic order and the existence of the ferromagnetic-ferromagnetic
transition.  This indicates that the time-scale order plays a
significant role in phase transitions and cooperative phenomena. 
An interesting and open problem would be to see if the results found in
the spin-lattice gas model are preserved beyond the mean-field analysis,
for instance in finite-dimensional short range models. In this
direction, we further progress for the model up to the Bethe approximation\cite{NH2}.

\begin{acknowledgments}
This work was supported by the Grant-in-Aid Scientific Research on the
 Priority Area ``Deeping and Expansion of  Statistical Mechanical
 Informatics'' (No. 1807004) by Ministry of Education, Culture, Sports,
 Science and Technology, Japan. 
\end{acknowledgments}

\end{document}